\documentclass[article]{revtex4-1}
\usepackage{amsmath}
\usepackage{mathrsfs} 
\usepackage{amssymb} 
\usepackage{xcolor,graphicx}
\usepackage{subfigure}
\usepackage{dcolumn}
\usepackage{bm}
\usepackage{natbib}

\begin{document}

\title{Distinct spin-lattice and spin-phonon interactions in monolayer magnetic CrI$_3$}

\author{Lucas Webster}
\affiliation{Department of Physics, Astronomy, and Geosciences, Towson University, 8000 York Road, Towson, MD 21252, USA}
\author{Liangbo Liang}
\affiliation{Center for Nanophase Materials Sciences, Oak Ridge National Laboratory, Oak Ridge, Tennessee 37831, USA}
\author{Jia-An Yan}
\email{jiaanyan@gmail.com}
\affiliation{~~Department of Physics, Astronomy, and Geosciences, Towson University, 8000 York Road, Towson, MD 21252, USA}





\begin{abstract}
 We apply the density-functional theory to study various phases (including non-magnetic (NM), anti-ferromagnetic (AFM), and ferromagnetic (FM)) in monolayer magnetic chromium triiodide (CrI$_3$), a recently fabricated 2D magnetic material. It is found that: (1) the introduction of magnetism in monolayer CrI$_3$ gives rise to metal-to-semiconductor transition; (2) the electronic band topologies as well as the nature of direct and indirect band gaps in either AFM or FM phases exhibit delicate dependence on the magnetic ordering and spin-orbit coupling; and (3) the phonon modes involving Cr atoms are particularly sensitive to the magnetic ordering, highlighting distinct spin-lattice and spin-phonon coupling in this magnet. First-principles simulations of the Raman spectra demonstrate that both frequencies and intensities of the Raman peaks strongly depend on the magnetic ordering. The polarization dependent $A_{1g}$ modes at 77 cm$^{-1}$ and 130 cm$^{-1}$ along with the $E_g$ mode at about 50 cm$^{-1}$ in the FM phase may offer a useful fingerprint to characterize this material. Our results not only provide a detailed guiding map for experimental characterization of CrI$_3$, but also reveal how the evolution of magnetism can be tracked by its lattice dynamics and Raman response.
\end{abstract}


\maketitle

\section{Introduction\label{intro}}

The discovery of intrinsic ferromagnetism in monolayers of CrGeTe$_3$ \cite{Gong2017} and CrI$_3$ \cite{Huang2017} has sparked tremendous interest in magnetism in the two-dimensional (2D) monolayer limit. Combining ferromagnetism in 2D materials is highly desirable not only for studying the fundamental physics of magnetism in low dimensions, but also for expanding the possibilities for new technological applications such as nanoscale spintronics, which aims to exploit the spin degree of freedom of electrons to carry information \cite{Prinz1998}, as an alternative to conventional charge based electronics. Linear magnetoelectric effect has recently been demonstrated in devices fabricated from bilayer CrI$_3$ \cite{Jiang2018,Huang22017,ShengweiJiang2018}. Combining CrI$_3$ and WSe$_2$ monolayers enabled unprecedented control over spin/valley physics \cite{Zhong2017}. On the basis of first principle calculations, Chern insulating state is predicted in CrI$_3$/graphene \cite{Zhang2017} and CrI$_3$/Bi$_2$Se$_3$/CrI$_3$ \cite{Hou2018} heteroestructures. The 2D magnets also render an exciting platform for studying the interplay between light and magnetic ordering \cite{Seyler2017}.

Raman spectroscopy has been widely employed to characterizing materials and probing the electronic properties and lattice dynamics of 2D materials \cite{ZhangX2016}. In principle, exchange coupling between magnetic ions can affect Raman response, and the coupling of phonons with multiferroic properties have been extensively studied in hexagonal manganites \cite{Flores2006,Litvinchuk2004,Souchkov2002}, perovskite oxides \cite{Haumont2006} and iron chalchogenides \cite{Opacic2017,Wang2016,Zhang2010}. Recently, Wang \emph{et al.} detected spin-order induced Raman peak at N\'{e}el temperature of FePS$_3$\cite{Wang2016}. Such behavior persists down to the monolayer limit, indicating strong in-plane spin-phonon coupling \cite{Wang2016}. Moreover, it has been shown that Raman spectroscopy can be used to probe the magnetic phase transition temperature \cite{Wang2016}.

In light of these facts, a better knowledge of spin-phonon coupling will enhance our understanding and facilitate possible applications of 2D magnetic materials. It is thus crucial to investigate effects of different magnetic orderings on the electronic and vibrational properties, but such a comprehensive investigation lacks for CrI$_3$, since the research of this interesting material is still in the early stage. Here, we carried out a systematic theoretical study on how spin affects the electronic structures, lattice dynamics, and Raman response in monolayer CrI$_3$. Various spin configurations, including nonmagnetic (NM), ferromagnetic (FM) and anti-ferromagnetic (AFM) orderings, are considered. Based on first-principles density functional theory (DFT) calculations, we found that the introduction of magnetism in monolayer CrI$_3$ gives rise to metal-to-semiconductor transition. The electronic band topologies as well as the nature of direct and indirect band gaps in either FM or AFM phases exhibit delicate dependence on the magnetic ordering and spin-orbit coupling (SOC).

Furthermore, our calculations reveal significant spin-lattice and spin-phonon coupling as different magnetic orderings lead to different lattice constants. For the phonon modes involving Cr atoms that host the magnetic moments of the system, they are found sensitive to the magnetic ordering. First-principles simulations of Raman spectra also show that both frequencies and intensities of Raman modes correlate with the magnetic ordering, demonstrating a strong dependence of both lattice dynamics and Raman response on the magnetic ordering in this 2D magnet. Our results not only provide a detailed guiding map for experimental characterization of CrI$_3$, but also reveal how the evolution of magnetism (with temperature or other factors) might be tracked by its lattice dynamics and Raman response. The investigation of a magnetic 2D crystal with different magnetic ordering is of great importance to the field as growing attention has been paid to the magnetism in two-dimensions.

\section{Calculational Methods \label{method}}

\begin{figure}[tbp]
\centering
\includegraphics[height=8.5 cm,angle=-90, clip]{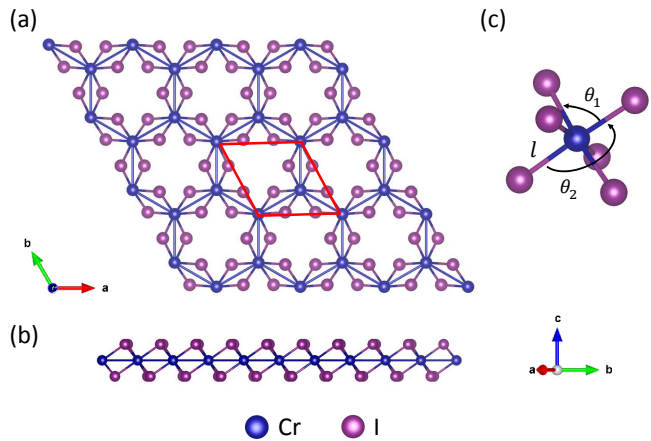}
\caption{\label{fig:model}Atomic structure of monolayer CrI$_3$. (a) Top view and (b) side view of monolayer CrI$_3$; (c) Bonding between chromium and iodine atoms. The unit cell of CrI$_3$ which includes two Cr and six I atoms has been indicated in (a). The bond length $l$ between Cr and I atom, the bond angle $\theta_1$ between Cr and two I atoms in the same plane, and the axial angle $\theta_2$ are also shown in (c). }
\end{figure}

We performed DFT calculations using projected augmented wave (PAW) method as implemented in the Vienna ab initio Simulation Package (VASP) \cite{Kresse1996,Kresse1999}.  To provide a gauge of how the lattice dynamics in this magnetic system depends on the exchange correlation (XC) functional, we have carried out calculations using both local density approximation (LDA) in Perdew-Zunger scheme \cite{Perdew1981} and generalized-gradient approximation with Perdew-Burke-Ernzerhof (PBE) \cite{Perdew1996}. It is found that overall the phonon frequencies and Raman spectra calculated in LDA yield better agreement with experimental data. Hence, we will focus on the LDA data in the main text, leaving the PBE results in the Supporting Materials for readers' reference.

The spin-polarized calculations were carried out to reveal the effects of magnetic ordering on the structural and electronic properties of this material. Two different magnetic orderings were considered: The ferromagnetic (FM) configuration had all magnetic moments initialized in the same direction while in the antiferromagnetic (AFM) configuration the magnetic moments were set to be antiparallel between two Cr atoms in the unit cell. For calculations including SOC, spin orientations were initialized along the out-of-plane $z$ direction. For comparison, the non-spin-polarized nonmagnetic (NM) phase has also been considered. The cut-off energy of the plane wave expansions was set to be 500 eV. The Brillouin Zone (BZ) was sampled using $\Gamma$-centered 4$\times$4$\times$1 uniform $k$-grid. A test of denser 8$\times$8$\times$1 $k$-grid yields nearly unchanged results for both electronic structures and lattice dynamics. The energy convergence value between two consecutive steps was chosen as 10$^{-6}$ eV. For monolayer CrI$_3$, a large vacuum of more than 20 \AA~ was applied along the perpendicular $z$ direction to avoid artificial interactions between images. Both unit cell and atomic positions have been fully relaxed, with the force converged to be below 2$\times$10$^{-3}$ eV/\AA. For the lattice dynamics, we used Phonopy package \cite{Togo2015}. The Raman calculations were carried out using the methods developed in Ref. \cite{Liang2014}. Since CrI$_3$ in the NM state is metallic, the dielectric tensor for Raman calculations was obtained at a typical experimental laser frequency 1.96 eV (633 nm). Using the finite dynamic dielectric tensor at the laser frequency is physically correct in the Placzek approximation \cite{Profeta2001,delCorro2016,Amos2000,Heyen1990}.

\section{Results and Discussions}

\subsection{Atomic structures}

The bulk chromium triiodide CrI$_3$ is a layered van der Waals material and can be mechanically exfoliated to produce 2D monolayers \cite{Huang2017}. This compound undergoes a structural phase transition from monoclinic AlCl$_3$ structure (space group $C2/m$) to rhombohedral BiI$_3$ structure (space group $R\bar{3}$) upon cooling \cite{McGuire2015,McGuire2017}. In both phases, the chromium ions form a honeycomb network sandwiched by two atomic planes of iodine atoms as shown in Figs.~\ref{fig:model}(a) and ~\ref{fig:model}(b). The parallelogram in Fig.~\ref{fig:model}(a) highlights the unit cell which contains two chromium and six iodine atoms per layer. Moreover, Cr$^{3+}$ ions are coordinated by edge-sharing octahedra, as shown in Fig.~\ref{fig:model}(c).

Magnetism in this compound is associated with the partially filled $d$ orbitals, as Cr$^{3+}$ ion has an electronic configuration of 3$d^{3}$. In the octahedral environment, crystal field interaction with the iodine ligands results in the quenching of orbital moment ($L$ = 0) and splitting of the chromium $d$ orbitals into a set of three lower energy $t_{2g}$ orbitals, and two higher energy $e_g$ orbitals. Therefore, according to Hund's rule, the three electrons occupying the $t_{2g}$ triplet will have $S$ = 3/2, which gives an atomic magnetic moment of 3 $\mu_B$. This picture is consistent with the observed saturation magnetization of bulk CrI$_3$ \cite{Richter2018}.

According to experiments, CrI$_3$ is a ferromagnetic semiconductor with the Curie temperature T$_c$ $\sim$ 61-68 K \cite{McGuire2015,McGuire2017}. Interestingly, CrI$_3$ is expected to retain its ferromagnetism when mechanically exfoliated down to the single layer limit, and the Curie temperature of the single layer is found to be T$_c$ $\sim$ 45 K \cite{Huang2017}. The magnetic moments lie along the $c$ lattice vector (perpendicular to the basal plane) in both bulk and single layer CrI$_3$.

The calculated structural parameters for different magnetic phases (NM, AFM, and FM) are shown in Table~\ref{tab1}. The available experimental results for the bulk structure are also listed for comparison. As shown in Table~\ref{tab1}, the total energy $E_t$ per unit cell shows the relative stability of CrI$_3$ in different phases with and without SOC. Clearly, the FM phase is energetically more favorable than NM and AFM and is thus the ground state, in agreement with experiment. Specifically, the total energy in FM is 14 meV lower than that of the AFM phase. The inclusion of SOC changes this difference by only 1 meV. As can be seen from Table S1 in the Supporting Documents, PBE also confirms that FM is the ground state with much larger energy difference (35 meV). Further LDA+U and PBE+U calculations with U = 3 eV \cite{YangK2009} also show that the FM phase is the ground state for monolayer CrI$_3$ (see Table S2 in the Supporting Document). In the following, we will compare the results between NM, AFM and FM to understand how the magnetic ordering affects the properties in monolayer CrI$_3$.

The obtained lattice constants for each phase are also listed in Table~\ref{tab1}. Compare with the experimental lattice constant (6.867 \AA) for the bulk, the lattice constant for the monolayer in LDA is about 0.2 \AA~ (nearly 2-3\%) smaller in all phases, which is common in LDA calculations. We find that the lattice constant in the FM phase increases very slightly ($\sim$ 0.003 \AA) after inclusion of SOC. In contrast, the Cr-I bond length in the NM phase is 2.602 \AA, about 4.5\% smaller than the experimental value. After inclusion of magnetization, the value increases to about 2.66 \AA, within 2.4\% of the experimental data. This result demonstrates the effect of spin on the atomic structure in CrI$_3$. The bond angle $\theta_1$ formed between I-Cr-I atoms is approximately 90$^\circ$ in both NM and AFM phases (this angle decreases by more than 8$^\circ$ in the NM phase as predicted in PBE, see Supporting Documents), while about 3$^\circ$ smaller in the FM phase. The so-called axial angle $\theta_2$ is the angle formed between the chromium ion and two opposing ligands within the same octahedral. From Table~\ref{tab1}, it turns out that this angle is smaller than 180$^\circ$.

\begin{table*}[tbp]
\centering
 \caption{Total energy $E_t$ (in eV/cell), optimized lattice constant $a_0$ (in \AA), bond length $l$ (in \AA), bond angle $\theta_1$ and $\theta_2$, and the energy band gap $E_g$ (in eV) for monolayer CrI$_3$ in NM, AFM and FM phases under LDA with and without SOC, respectively. Experimental data for the bulk are listed for comparison.} \label{tab1}
\begin{tabular}{lcccccccccc}
\hline
   & \multicolumn{1}{c}{NM} & \multicolumn{2}{c}{AFM} & \multicolumn{2}{c}{FM} & Exp. \\
   \cline{2-2} \cline{3-4} \cline{5-6}
   & LDA & LDA & LDA-SOC & LDA  & LDA-SOC & bulk\cite{McGuire2015}\\
   \hline
   $E_t$ 	& -33.272&  -35.372    & -36.243           & -35.386 & -36.256 & \\
    $a_0$  	& 6.645 & 	6.667	        & 6.669         &	6.686	& 6.689 & 6.867\\
    $l$    	& 2.602 & 	2.653	        & 2.656         &	2.655	& 2.658	& 2.725\\
$\theta_1$ 	& 90.1$^\circ$ &	89.9$^\circ$   & 89.9$^\circ$  &	 86.7$^\circ$ &	90.2$^\circ$ & \\
$\theta_2$ 	& 172.8$^\circ$ &	175.4$^\circ$  & 175.5$^\circ$ & 175.6$^\circ$	& 175.7$^\circ$ & \\
    $E_g$  	& -  &	1.247          & 0.959         &	1.135 &	0.918 & 1.2 \\
   \hline
\end{tabular}
\end{table*}

\subsection{Electronic structures}

Previous DFT calculations based on PBE have been reported for chromium trihalides \cite{McGuire2015, Zhang2015, Wang2011}. However, investigations of the electronic properties in different magnetic phases and particularly the lattice dynamics in this material are still lacking. In the following, we will mainly focus on the results obtained from LDA.

\begin{figure*}[tbp]
\centering
\includegraphics[width=0.9\textwidth, clip]{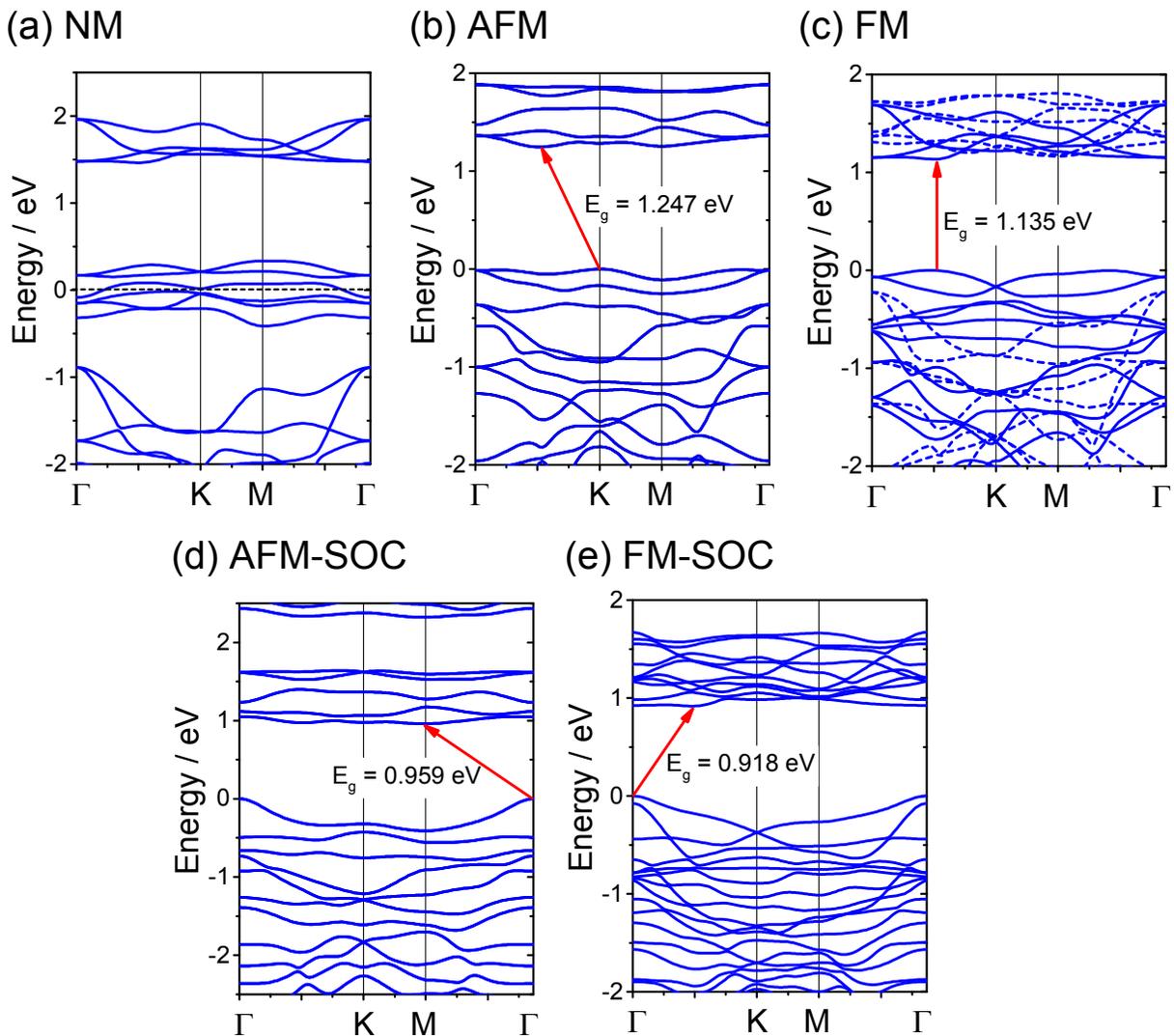}
\caption{\label{fig:band}Calculated electronic band dispersions of monolayer CrI$_3$ for the NM, AFM and FM phases, respectively, in LDA. (a) NM phase; (b) AFM phase without SOC; (c) FM phase without SOC; (d) AFM phase with SOC; (e) FM phase with SOC. In (b) and (c), solid lines denote spin-up while dashed lines denote spin-down bands. The energy band gaps between conduction band minimum (CBM) and valence band maximum (VBM) have been indicated using red arrows in each case. The Fermi level or the VBM has been shifted to zero.}
\end{figure*}

Shown in Figs.~\ref{fig:band}(a)-(c) are the electronic band structures for the NM, AFM and FM phases, respectively. Non-collinear results for the AFM and FM phases including SOC are plotted in Figs.~\ref{fig:band}(d) and ~\ref{fig:band}(e), respectively. Clearly, the NM phase exhibits zero band gap, in agreement with previous DFT calculations \cite{McGuire2015}. This is understandable: in the NM phase, the lowest $t_{2g}$ orbitals are half-filled with $3d^3$ electron configuration for the Cr$^{3+}$ ions, leading to metallic state. In contrast, the spin-polarized calculations in either AFM or FM phases predict a finite band gap near the Fermi level. This implies that a Mott-Hubbard mechanism plays a key role in the formation of the band gap in this material \cite{McGuire2015,Zhang2015}.

There are a few important features revealed in Fig.~\ref{fig:band}. First, the spin-up (solid) and spin-down (dashed) bands in the FM phase split in Fig.~\ref{fig:band}(c) near the band edges, where spin-up bands dominate, leading to a net magnetic moment ($\sim$ 3.0 $\mu$B/Cr). In contrast, the spin-up (solid) and spin-down (dashed) bands in the AFM phase are almost overlapped as shown in Fig.~\ref{fig:band}(b), which is understandable because of the zero net magnetic moment in the AFM phase. This finding agrees well with a very recent theoretical work by Larson \emph{et al.} \cite{Larson2018}. Secondly, comparing FM with AFM, we find that the VBM shifts from $K$ in AFM to the middle of the $\Gamma-K$ line in FM, while the location of the CBM remains the same, leading to an indirect-to-direct transition. Third, the band topologies near the VBM and CBM are also distinct between the AFM and FM phases. These results demonstrate that the magnetic ordering has important implications on the electronic band structures, especially at the band edges. More interestingly, the inclusions of the SOC in the AFM and FM phases completely change the corresponding band dispersions near the band edges, as shown in Figs.~\ref{fig:band}(d) and ~\ref{fig:band}(e), respectively. Specifically, the VBM now shifts from $K$ in Fig.~\ref{fig:band}(b) to $\Gamma$ in Fig.~\ref{fig:band}(d), while the CBM changes to $M$. The degeneracy at $\Gamma$ is also lifted due to SOC. In contrast, in the FM phase, only the VBM shifts to the $\Gamma$ after inclusion of SOC, resulting in a direct-to-indirect transition on the band gap. Apparently, distinct from the transition metal dichalcogenides \cite{Yan2015}, the effects of SOC are crucial in determining the electronic band structures in this magnetic material. These results, however, are slightly different from the PBE data, as shown in Fig.~S1. Since the electronic band structures play a crucial role in many device applications, our results call for a further experimental (e.g., angle-resolved photoemission spectroscopy) measurement of the band dispersions, which may help to verify the band dispersions in this material.

In each case, the energy band gaps, ranging from 0.918 eV to 1.247 eV between the CBM and VBM, have been indicated in Fig.~\ref{fig:band}. The data are also listed in Table~\ref{tab1}. In contrast to the FM ground state, the AFM phase exhibits similar semiconducting nature with the band gap nearly 9.9\% (4.5\% with SOC) larger than those of the FM phases. Interestingly, the inclusion of SOC decreases the band gap dramatically: for the FM phase, the band gap in FM-SOC is 0.918 eV, decreasing by 0.217 eV from that of the FM phase, about 19.1\%. In the AFM phase, the band gap decreases by 23\% from 1.247 eV to 0.959 eV after including SOC. These results highlight the crucial role of SOC in determining not only the band topologies, but also the band gaps. The energy band gap (1.135 eV) in the FM phase is in close agreement with previous result (1.143 eV) reported in Ref. \cite{Zhang2015}.

\subsection{Lattice dynamics}

\begin{figure}[tbp]
\centering
\includegraphics[width=8.5cm, clip]{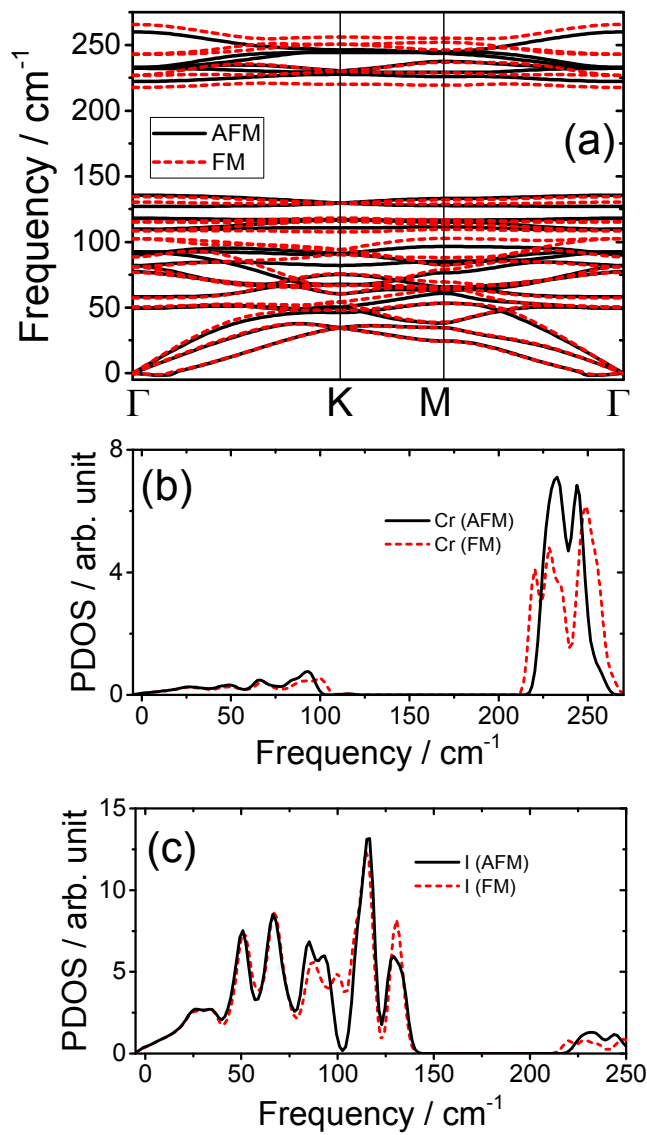}
\caption{\label{fig:phband}Calculated phonon band dispersions for monolayer CrI$_3$. (a) AFM (solid) and FM (dashed). The corresponding partial phonon density of states are shown for two Cr atoms (b) and six I atoms (c), respectively. }
\end{figure}

Our calculations in LDA show that there is a significant negative phonon branches in the NM phase, in contrast to the PBE data as shown in Fig.~S1. Hence, we will focus only on the AFM and FM data as calculated in LDA. The phonon band dispersions for the FM (dashed) and AFM (solid) phases are shown in Fig.~\ref{fig:phband}(a). A few important features emerge. First, for both phases, there are no negative phonon branches, indicating that both phases are mechanically stable (Tiny negative phonon frequencies are generally believed to be induced by calculation precision and will not be of importance). Second, as shown in Fig.~\ref{fig:phband}(a), a clear phonon band gap of about 85 cm$^{-1}$ ranging from 135 cm$^{-1}$ to 220 cm$^{-1}$ can be identified. Interestingly, the gap formation on the phonon band dispersion after including the magnetic ordering is more or less similar to the electronic case as shown in Fig.~\ref{fig:band} (It becomes more evident if we consider the PBE data, see Fig.~S2). Clearly, the phonon band gap here is also induced by the magnetic ordering. Finally, we observe a visible shift of the optical phonon branches between the AFM and FM phases. This feature suggests a strong spin-phonon coupling in this magnetic material. In contrast, the acoustic phonon branches are less sensitive to the magnetic ordering in Fig.~\ref{fig:phband}(a) since there is almost negligible difference on the acoustic branch between those of the AFM and FM phases.

The above observations can be further understood from the partial phonon density of states (PDOS) of the Cr and I atoms for both the AFM and NM phases, as depicted in Figs.~\ref{fig:phband}(b) and ~\ref{fig:phband}(c), respectively. The atomic mass of I atom is 126.904 u, while it is 51.996 u for the Cr atom. Due to the larger mass of the I atoms, the PDOSs of the low-frequency modes ($<$150 cm$^{-1}$) are mainly dominated by the I atoms, while the high-frequency modes ($>$200 cm$^{-1}$) mainly consist of vibrations of Cr atoms, which are smaller in atomic mass. As can be seen clearly from Fig.~\ref{fig:phband}(b), it turns out that the spin-phonon coupling is more evident in the high-frequency vibrational modes involving Cr atoms than other modes. This is understandable since Cr atoms carry most of the magnetic moments. The results explains why evident deviations can be observed for the optical phonon branches in the AFM and FM phases.

We now focus on the vibrational modes at $\Gamma$. This is of importance for understanding the Raman spectra in this material. The monolayer CrI$_3$ in the NM phase possesses $D_{3d}$ point group symmetry. The phonon modes at $\Gamma$ can be decomposed as $\Gamma_{D_{3d}}=2 A_{1g}+2 A_{2g}+4 E_g+2 A_{1u}+2 A_{2u}+4 E_u$. Excluding three acoustic modes (doubly degenerate $E_u$ + 1 $A_{1u}$), there are 21 modes in total. Since the ground state is FM, we will focus on the modes in this phase. Table~\ref{tab2} lists all the 21 phonon vibrational modes in the AFM and FM phases calculated using LDA, along with their group irreducible representations.

Several conclusions can be reached from Table~\ref{tab2}. First, the frequencies of the modes 1 and 2 ($E_g$), 3 ($A_{2u}$) and 4 ($A_{1g}$), 5 and 6 ($E_u$), 7 ($A_{2g}$), 12 and 13 ($E_u$), 15 ($A_{2u}$) are only changed slightly between AFM and FM phases, while all other modes exhibit distinct deviations between the two magnetic phases. For example, the highest $A_{1u}$ mode in the AFM is 259.0 cm$^{-1}$, nearly 5.7 cm$^{-1}$ lower than that of the FM phase. The $E_g$ mode at 93.0 cm$^{-1}$ in AFM is about 10 cm$^{-1}$ smaller than that of FM phase. Second, comparing the data between AFM and FM phases, we find that the deviations are more common for the high-frequency modes than the low-frequency modes. For instance, mode 1 is at 49.6 cm$^{-1}$ in AFM, comparable to 50.5 cm$^{-1}$ in the FM, while mode 16 ($A_{2g}$ mode) is 220.2 cm$^{-1}$ in AFM, nearly 2.6 cm$^{-1}$ larger than 217.6 cm$^{-1}$ in the FM phase. In particular, the high-frequency $E_g$ mode at 244.7 cm$^{-1}$ in FM is extremely sensitive to the spin and decreases by 10 cm$^{-1}$ in the AFM phase. Finally, the degenerate modes as calculated in LDA split into slightly different branches after inclusion of SOC. Nevertheless, the splittings on the mode frequencies are very small, within 0.5 cm$^{-1}$ in most cases. Thus, effects of SOC on the mode frequencies can be safely neglected in LDA.

It should be mentioned that the phonon frequencies at $\Gamma$ are highly sensitive to the XC functionals employed, especially the high-frequency modes involving magnetic element Cr. Comparing to the PBE data in Table S3 in the Supporting Documents, significant deviations can be identified for the high-frequency modes. For example, for the mode 19, the LDA yields a frequency of 244.7 cm$^{-1}$ in the FM phase, nearly 23.7 cm$^{-1}$ higher than that of PBE in the FM phase. In contrast, for the low-energy mode 4, the difference is only 9.5 cm$^{-1}$. Similar but less evident deviations can be identified for the AFM phase as well.

\begin{figure*}[tbp]
\centering
\includegraphics[width=0.75\textwidth, clip]{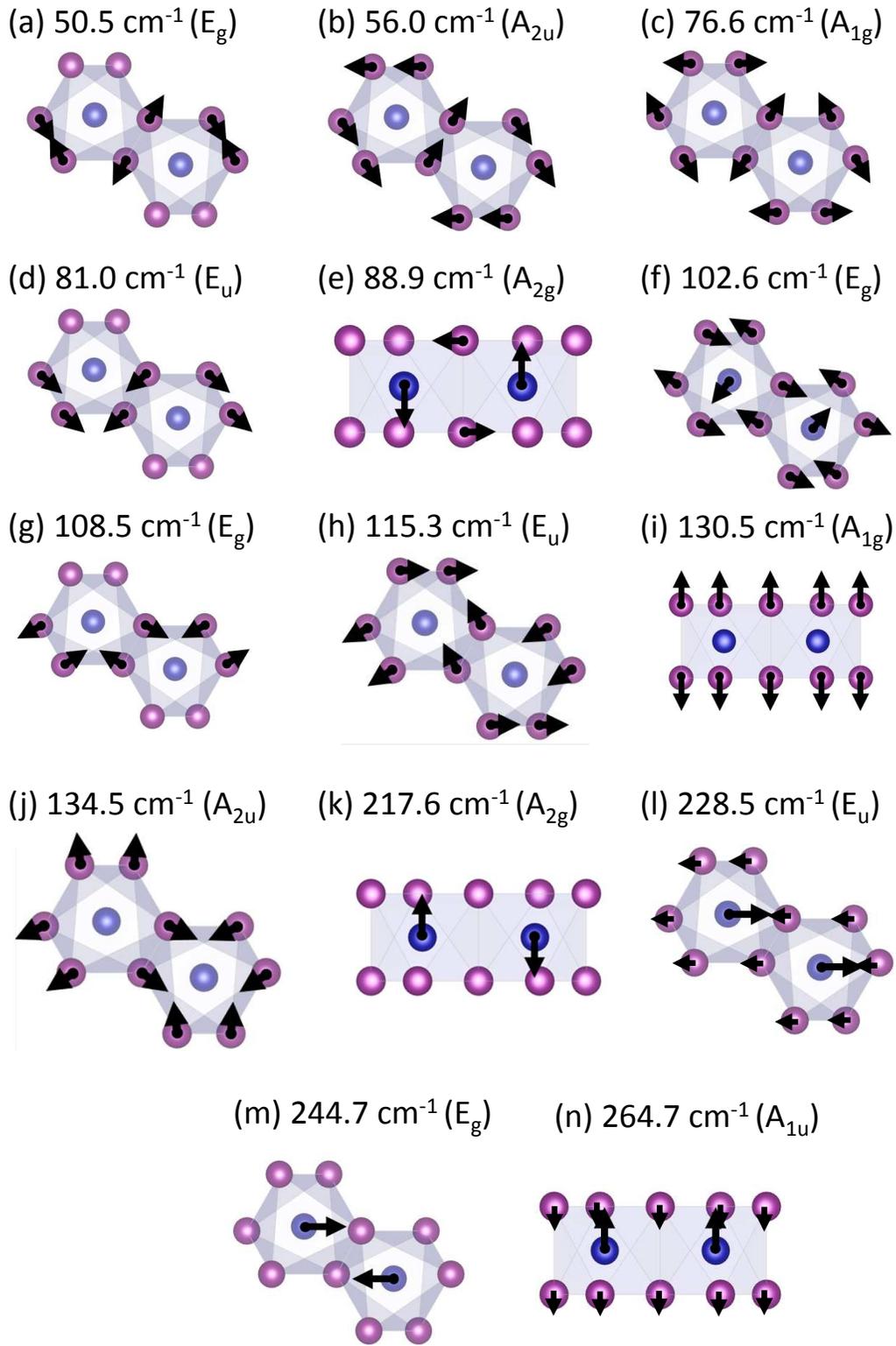}
\caption{\label{fig:mode}Schematic representations of the eigenvectors for all 14 distinct phonon vibrational modes at $q$ = $\Gamma$ in the FM phase calculated using LDA. The group representations for each mode in the FM phase are also listed for reference.}
\end{figure*}

The eigenvectors of these 21 modes have been schematically shown in Fig.~\ref{fig:mode}. For the degenerate cases, we only show one pattern. There are some interesting characteristics to be mentioned. The $A_{1u}$ mode at 264.7 cm$^{-1}$ shown in Fig.~\ref{fig:mode}(n) consists of out-of-plane vibrations between Cr and two I layers. This pattern in which each layer vibrates as a whole is more or less similar to the layer breathing modes in van der Waals 2D materials but with much higher frequency. The $E_g$ mode at 244.7 cm$^{-1}$ in Fig.~\ref{fig:mode}(m) mainly involves in-plane out-of-phase vibrations of two Cr atoms. Similarly, the $E_u$ mode at 228.5 cm$^{-1}$ in Fig.~\ref{fig:mode}(l) involves in-plane vibrations of two Cr atoms; however the vibrations are in-phase and they move relative to the I atoms. The $A_{2g}$ mode at 217.6 cm$^{-1}$ in Fig.~\ref{fig:mode}(k) is similar to the highest-frequency mode in Fig.~\ref{fig:mode}(n) as it also consists of out-of-plane vibrations, except that in this case the Cr atoms move out-of-phase within the unit cell. This mode is Raman inactive. Clearly from Fig.~\ref{fig:mode}(k) to Fig.~\ref{fig:mode}(n), these high-frequency vibrations mainly involve motions of Cr atoms since the atomic mass of Cr is much smaller than I. Also the vibrational amplitudes of Cr are much larger than those of the I atoms.

Below 200 cm$^{-1}$, the modes mainly arise from the I atomic vibrations, except the $A_{2g}$ mode at 88.9 cm$^{-1}$ in Fig.~\ref{fig:mode}(e), which is similar to the mode shown in Fig.~\ref{fig:mode}(k). However, in this case the I ions vibrate in-plane (120$^\circ$ out-of-phase) so that the bond length is kept nearly constant. The mode of $A_{2u}$ at 134.5 cm$^{-1}$ shows interesting patterns for the six I atoms, expanding and shrinking in-plane around the Cr atom, as shown in Fig.~\ref{fig:mode}(j). The $A_{1g}$ mode at 130.5 cm$^{-1}$ in Fig.~\ref{fig:mode}(i), which is Raman active, consists of the out-of-phase and out-of-plane vibrations of two I layers. The modes shown in Figs.~\ref{fig:mode}(f), (d) and (a) are similar in the sense that in these modes the armchair chain of I atoms forms a vibrating strip.

The $E_g$ mode portrayed in Fig.~\ref{fig:mode}(f) resembles a shearing motion of the iodine planes. However, small out-of-plane displacements can also be found (not shown in the picture). The $A_{1g}$ mode at 76.6 cm$^{-1}$ shown in Fig.~\ref{fig:mode}(c) and the $E_g$ mode at 108.5 cm$^{-1}$ shown in Fig.~\ref{fig:mode}(g) also display the dominance of the vibrations of the paired I atoms out of phase. These modes are Raman active. Finally, the $A_{2u}$ mode shown in Fig.~\ref{fig:mode}(b) involves a small rotation of the I atoms around the central Cr atom.

Among all the 21 modes, the $E_g$, $A_{1g}$ and $A_{2g}$ modes are Raman active from symmetry point of view. However, the Raman intensities of the $A_{2g}$ mode are negligible.

\begin{table*}[h]
\centering
 \caption{Frequencies of the phonon vibrational modes for monolayer CrI$_3$ at $\Gamma$ calculated under LDA with and without SOC for the AFM and FM phases. The Raman active modes have been highlighted in bold. The unit is in cm$^{-1}$.} \label{tab2}
\begin{tabular}{llllllllc}
\hline
& \multicolumn{2}{c}{AFM} & \multicolumn{2}{c}{FM} &   \\
    \cline{2-3} \cline{4-5}
Mode No.      & LDA & LDA-SOC & LDA  & LDA-SOC & Symmetry\\
   \hline
1, 2 	&  \textbf{49.6}  &	\textbf{48.9, 49.1} 	& \textbf{50.5} & \textbf{ 50.1, 50.2 } & \textbf{$E_g$}  \\
3	 	&  58.4  &	58.1	  	& 56.0 &  56.8	   	  & $A_{2u}$ \\
4	 	&  \textbf{77.2}  & \textbf{76.8}	    & \textbf{76.6} &  \textbf{76.1}		  &   \textbf{$A_{1g}$} \\
5, 6 	& 82.8   & 82.2, 82.4 & 81.0 &  80.3, 80.5  & $E_u$ \\
7	 	&  90.2  & 90.0		& 88.9 &  88.0		  &  $A_{2g}$ \\
8, 9 	& \textbf{93.0} & \textbf{91.7, 91.9}	& \textbf{102.6}& \textbf{101.8, 101.9} &  \textbf{$E_g$} \\
10, 11 	& \textbf{110.0}	& \textbf{109.0, 109.1}& \textbf{108.5} & \textbf{107.5, 107.6}	& \textbf{$E_g$ }\\
12, 13  &  117.7	& 116.6, 116.7& 115.3	& 114.3, 114.3	&	$E_u$ \\
14	   	&  \textbf{127.0}	& \textbf{125.7}		  & \textbf{130.5}	& \textbf{129.0}			& \textbf{$A_{1g}$ }\\
15	   	&  135.5	& 134.3       & 134.5	& 133.3			& $A_{2u}$ \\
16	   	&  220.2	& 218.4   	  & 217.6	& 215.5			& $A_{2g}$ \\
17, 18  &  234.3	& 230.8, 231.2& 228.5	& 225.3, 225.7  & $E_u$ \\
19, 20  &  \textbf{235.0}	& \textbf{232.3, 232.5}& \textbf{244.7}	& \textbf{241.1, 241.3}	& \textbf{$E_g$}\\
21	   	&  259.0	& 256.9		  & 264.7	& 262.3			& $A_{1u}$ \\
\hline
\end{tabular}
\end{table*}

\subsection{Raman spectra}

Raman spectroscopy is a powerful tool to probe various physical properties in materials. In particular, polarized Raman spectroscopy has been applied to investigate 2D materials, including crystalline orientations in ReS$_2$ and strained graphene \cite{Chenet2015,He2016,HuangM2009}, anisotropy of black phosphorus \cite{LingXi2015}, and dichroism of helical change in the light in MoS$_2$ \cite{ChenSY2015}. In these materials, the Raman intensities of specific modes are sensitive to the crystalline anisotropy, which can be probed through a certain laser polarization setup, i.e., the relationship between polarizations of the incident and scattered laser lights. Since the magnetic ordering dramatically changes the phonon vibrational modes as discussed above, we expect that (polarized) Raman spectra may also be employed to probe the magnetic properties in CrI$_3$.

\begin{figure*}[tbp]
\centering
\includegraphics[width=0.75\textwidth]{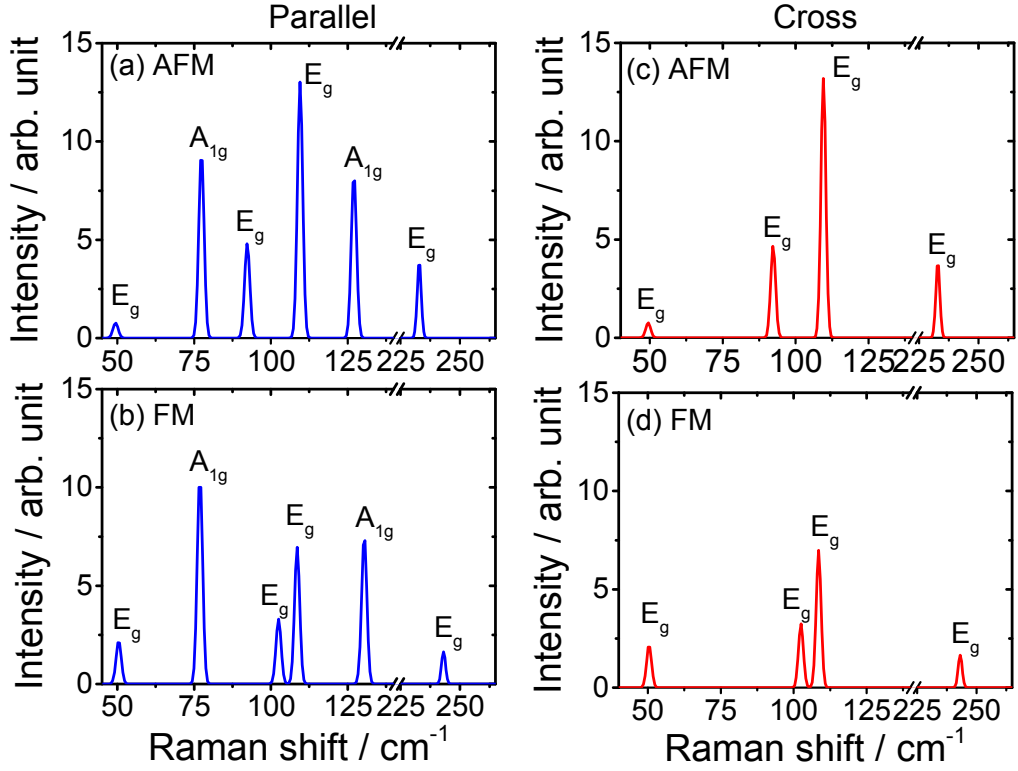}
\caption{\label{fig:raman}Polarized Raman spectra for monolayer CrI$_3$ in AFM and FM phases, respectively. Shown on the left are for the parallel polarization laser set-up. On the right are the results with cross laser polarization set-ups. The corresponding mode symmetries for each peak have been indicated. A Gausian broadening with width of 2.0 cm$^{-1}$ has been applied.}
\end{figure*}

To investigate how the magnetic ordering affects the Raman response, we have simulated the Raman spectra for monolayer CrI$_3$ in NM, AFM, and FM phases. Since LDA yields significant phonon branches with negative phonon frequencies, we will focus on the AFM and FM phases here (The PBE data for all NM, AFM and FM phases can be found in the Supporting Documents). We consider two typical polarizations, namely, parallel and cross, in the back-scattering laser set-ups that are common in experiment. Figs.~\ref{fig:raman}(a)-(b) show the results with parallel polarization for the AFM and FM phases, respectively, while Figs.~\ref{fig:raman}(c)-(d) correspond to those of cross polarization.

As shown in Figs.~\ref{fig:raman}(a) and ~\ref{fig:raman}(b), switching on the spin polarization leads to distinct Raman spectra in these magnetic phases. Not only the peak positions, but also the peak intensities change dramatically. In the AFM phase, these peaks are four $E_g$ modes at 49.6, 93.0, 110.0, and 235.0 cm$^{-1}$, and two $A_{1g}$ modes at 77.2 and 127.0 cm$^{-1}$, respectively. In the FM phase, the six Raman peaks, which should be readily observable in experiment, are four $E_g$ modes at 50.5, 102.6, 108.5 and 244.7 cm$^{-1}$, and two $A_{1g}$ modes at 76.6 and 130.5 cm$^{-1}$, respectively. The vibrational patterns of these modes have been illustrated in Figs.~\ref{fig:mode}(a), ~\ref{fig:mode}(f), ~\ref{fig:mode}(g),  ~\ref{fig:mode}(m), ~\ref{fig:mode}(c), and ~\ref{fig:mode}(i). Note that the peak positions in the FM phase also shift significantly relative to those of the AFM phase. Especially, the two $E_g$ peaks at 93.0 and 110.0 cm$^{-1}$ in the AFM phase evolve to two peaks with much smaller spacing ($\sim$ 6 cm$^{-1}$) in the FM phase.

Finally, in contrast with the PBE data in which the Raman peak above 200 cm$^{-1}$ is small (see Fig. S3), the $E_g$ peak at about 240 cm$^{-1}$ is clearly present in LDA, in good agreement with experimental observation \cite{Larson2018}.

Figs.~\ref{fig:raman}(c) and ~\ref{fig:raman}(d) illustrate the Raman spectra with cross polarization for the AFM and FM phases, respectively. Specifically, the two $A_{1g}$ Raman peaks at 77.2 and 127.0 cm$^{-1}$ in Fig.~\ref{fig:raman}(c) disappear in Fig.~\ref{fig:raman}(d). Similar features can be found for the FM phase as well. Since the two phases exhibit similar polarization dependence, we conclude that the polarization Raman spectra are only determined by the mode symmetry. The magnetic ordering affects the peak positions and intensities, but will not change the polarization dependence.

The dramatic distinctions between the FM and AFM phases demonstrate the coupling between magnetic ordering and the lattice dynamics as well as their effects on the Raman response. These findings may serve as a detailed guiding map for experimental characterization of CrI$_3$. Furthermore, the evolution of the Raman spectra with respect to the different magnetic ordering demonstrates that the magnetic ordering with respect to temperature and/or other factors might be possibly tracked by the lattice dynamics in CrI$_3$, particularly through the Raman response.

The behavior of the $E_g$ and $A_{1g}$ modes can be understood from their Raman tensor. The Raman intensity of the $j$-th phonon mode reads as\cite{Liang2014,Danna2015}:
\begin{equation}
I(j) \propto |\hat{g}_s \cdot \overline{\overline{\alpha}}(j)\cdot \hat{g}_i^T|^2,
\end{equation}
where $\hat{g}_i$ and $\hat{g}_s$ are the polarization unit vectors of the incoming and scattered lights, respectively. The Raman susceptibility $\overline{\overline{\alpha}}(j)$ is a symmetric $(3\times3)$ tensor associated with the $j$-th phonon mode.

Our DFT calculations show that the Raman tensor of the $A_{1g}$ mode takes the following form:
\begin{equation} \label{eq1}
\overline{\overline{\alpha}}({A_{1g}})=\left(
\begin{array}{ccc}
    a & 0 & 0 \\
    0 & a  & 0 \\
    0 & 0 & b \\
  \end{array}
\right),
\end{equation}
while the Raman tensor of the $E_g$ mode is:
\begin{equation} \label{eq2}
\overline{\overline{\alpha}}({E_g})=\left(
\begin{array}{ccc}
    a' & c & 0 \\
    c & b'  & 0 \\
    0 & 0 & 0 \\
  \end{array}
\right).
\end{equation}
These numerical results are in agreement with the analysis from group theory \cite{Loudon1964}. For the parallel polarization, $\hat{g}_i=[1 0 0]$ and $\hat{g}_s=[1 0 0]$, the intensity of the $E_g$ and $A_{1g}$ modes can be expressed as:

\begin{eqnarray}
I(E_g)\propto a'^2,
\end{eqnarray}
and
\begin{eqnarray}
I(A_{1g})\propto a^2.
\end{eqnarray}
These elements of $a$ and $a'$ are usually not zero. Hence the corresponding Raman peaks for these modes are visible. In contrast, for the cross polarization, $\hat{g}_i=[1 0 0]$ and $\hat{g}_s=[0 1 0]$, the expression for the intensities of the $E_g$ and $A_{1g}$ mode in the cross polarization follows:
\begin{eqnarray}
I(E_g)\propto c^2,
\end{eqnarray}
while
\begin{eqnarray}
I(A_{1g})= 0.
\end{eqnarray}
In other words, as long as $c$ element of the Raman tensor of the $E_g$ mode is nonzero, there will be visible Raman peaks corresponding to these $E_g$ modes in the cross polarization. In contrast, the intensity for the $A_{1g}$ modes will become zero. This explains what we have seen in Fig.~\ref{fig:raman}.

\section{Conclusions}

In summary, we have studied the effects of magnetic ordering on the atomic and electronic structures, lattice dynamics and Raman response in monoalayer CrI$_3$ using density-functional theory. Three different phases including non-magnetic, anti-ferromagnetic, and ferromagnetic have been investigated.

While non-magnetic phase exhibits nearly zero band gap, the spin-polarized calculations in either AFM or FM phases predict a semiconducting nature of the monolayer CrI$_3$, with the band gap in the range of 0.9 to 1.3 eV. In the FM and AFM phases, the band topologies near the CBM and VBM as well as the direct and indirect nature of these band gaps exhibit a delicate dependence on the magnetic ordering and spin-orbit coupling, as predicted in both LDA and PBE calculations. These results demonstrate the crucial role of the magnetic ordering and spin-orbit coupling in determining the electronic structures of this magnetic system.

Furthermore, the lattice dynamics, especially the high-frequency phonon modes involving Cr atoms, are found sensitive to the magnetic ordering, in contrast to the acoustic branches that are nearly unchanged with respect to the magnetic ordering. These results highlight the spin-lattice and spin-phonon couplings in this magnetic material.

Our calculated Raman spectra show that some Raman modes are sensitive to the magnetic ordering. Specifically, the two $E_g$ modes at around 102 and 108 cm$^{-1}$ and the high-frequency $E_g$ mode at 244.7 cm$^{-1}$ in the FM phase might be useful to track the magnetic phase transition, for example, with respect to the temperature or external electric field. The distinct behavior of the two $A_{1g}$ modes at 77 and 130 cm$^{-1}$ together with the $E_g$ mode at 50 cm$^{-1}$ may offer a fingerprint to characterize this 2D material.

Finally, it should be pointed out that the physical properties of monolayer CrI$_3$, including electronic band dispersions, lattice dynamics and Raman spectra, are strongly dependent on the XC functionals. Further experimental studies such as ARPES  measurements will help to elucidate the band structures in this magnetic material.

\section{Conflicts of interest}

There are no conflicts of interest to declare.

\section*{Acknowledgements}
We thank Drs. Joshua C. H. Lui and Rui He for sharing their experimental Raman data of few-layer CrI$_3$ prior to publication. This work used the Extreme Science and Engineering Discovery Environment (XSEDE) Comet at the SDSC through allocation TG-DMR160101 and TG-DMR160088. L.W. and J.A.Y. acknowledge support from the NSF grant DMR 1709781 and support from the Fisher General Endowment and SET grants from the Jess and Mildred Fisher College of Science and Mathematics at Towson University. A portion of this research (Raman scattering modeling) used the computational package developed at the Center for Nanophase Materials Sciences, which is a DOE Office of Science User Facility operated by the Oak Ridge National Laboratory. L. L. was supported by Eugene P. Wigner Fellowship at the Oak Ridge National Laboratory and by the Center for Nanophase Materials Sciences.



\pagebreak
\appendix{
\textbf{\large Supporting Documents: Distinct spin-lattice and spin-phonon interactions in monolayer magnetic CrI$_3$}
\setcounter{equation}{0}
\setcounter{figure}{0}
\setcounter{table}{0}
\setcounter{page}{1}
\makeatletter
\renewcommand{\theequation}{S\arabic{equation}}
\renewcommand{\thetable}{S\arabic{table}}
\renewcommand{\thefigure}{S\arabic{figure}}
\renewcommand{\bibnumfmt}[1]{[S#1]}
\renewcommand{\citenumfont}[1]{S#1}

\subsection{Atomic structures in PBE}

Structural parameters calculated using PBE are listed in Table~\ref{supptab1}.

\begin{table}[h]
\centering
 \caption{Total energy $E_t$ (in eV/cell), optimized lattice constant $a_0$ (in \AA), bond length $l$ (in \AA), bond angle $\theta_1$ and $\theta_2$, and the energy band gap $E_g$ (in eV) for monolayer CrI$_3$ in NM, AFM, and FM phases calculated using PBE with and without SOC. } \label{supptab1}
\begin{ruledtabular}
\begin{tabular}{lcccccc}
   & NM  & \multicolumn{2}{c}{AFM} & \multicolumn{2}{c}{FM} & Exp.  \\
   \cline{2-2} \cline{3-4} \cline{5-6}
   & PBE & PBE & PBE-SOC & PBE & PBE-SOC & bulk \cite{McGuire2015}\\
   \hline
   $E_t$ &   -28.377 & -31.453  & -32.282       &  -31.488 & -32.319 & \\
    $a_0$  &  6.988  &	6.999         &	7.007   &	 6.999	& 7.008 & 6.867 \\
    $l$    &  2.670  &	2.734         & 2.738   &   2.735	& 2.740 & 2.725 \\
$\theta_1$ & 81.86$^\circ$  &	90.4$^\circ$  &	90.4$^\circ$  &	 90.6$^\circ$ &	90.6$^\circ$ & \\
$\theta_2$ & 169.2$^\circ$ &	172.8$^\circ$ & 172.8$^\circ$ &	173.2$^\circ$ &	173.3$^\circ$ & \\
    $E_g$  &    -     &	1.263         & 1.024         &	1.132 &	0.890 & 1.2 \\
\end{tabular}
\end{ruledtabular}
\end{table}

Results from both LDA+U and PBE+U calculations (using U = 3 eV for Cr \cite{YangK2009}) for monolayer CrI$_3$ are listed in Table~\ref{supptab2} below.

\begin{table}[h]
\centering
 \caption{Total energy $E_t$ (in eV/cell), optimized lattice constant $a_0$ (in \AA), bond length $l$ (in \AA), bond angle $\theta_1$ and $\theta_2$, and the energy band gap $E_g$ (in eV) for monolayer CrI$_3$ in NM, AFM, and FM phases calculated using PBE with and without SOC. } \label{supptab2}
\begin{ruledtabular}
\begin{tabular}{lcccc}
   & \multicolumn{2}{c}{AFM} & \multicolumn{2}{c}{FM} \\
   \cline{2-3} \cline{4-5}
   & LDA+U & PBE+U &    LDA+U    & PBE+U  \\
   \hline
   $E_t$ &   -31.597 & -28.051  & -31.635 & -28.109\\
   $a_0$ &     6.74  &	7.07    &  6.75   &  7.08 \\
   $l$   &     2.689  &	2.774  &  2.689  & 2.775 \\
$\theta_1$ & 87.4$^\circ$  & 90.1$^\circ$  & 87.2$^\circ$  &	90.5$^\circ$ \\
$\theta_2$ & 175.8$^\circ$ & 173.2$^\circ$ & 176.0$^\circ$ &	173.5$^\circ$ \\
\end{tabular}
\end{ruledtabular}
\end{table}

\subsection{Electronic band dispersions in PBE}
The electronic band dispersions for the NM, AFM, and FM phases calculated from PBE are shown in Fig.~\ref{fig:suppband}.
\begin{figure}[h]
\centering
\includegraphics[width=0.75\textwidth]{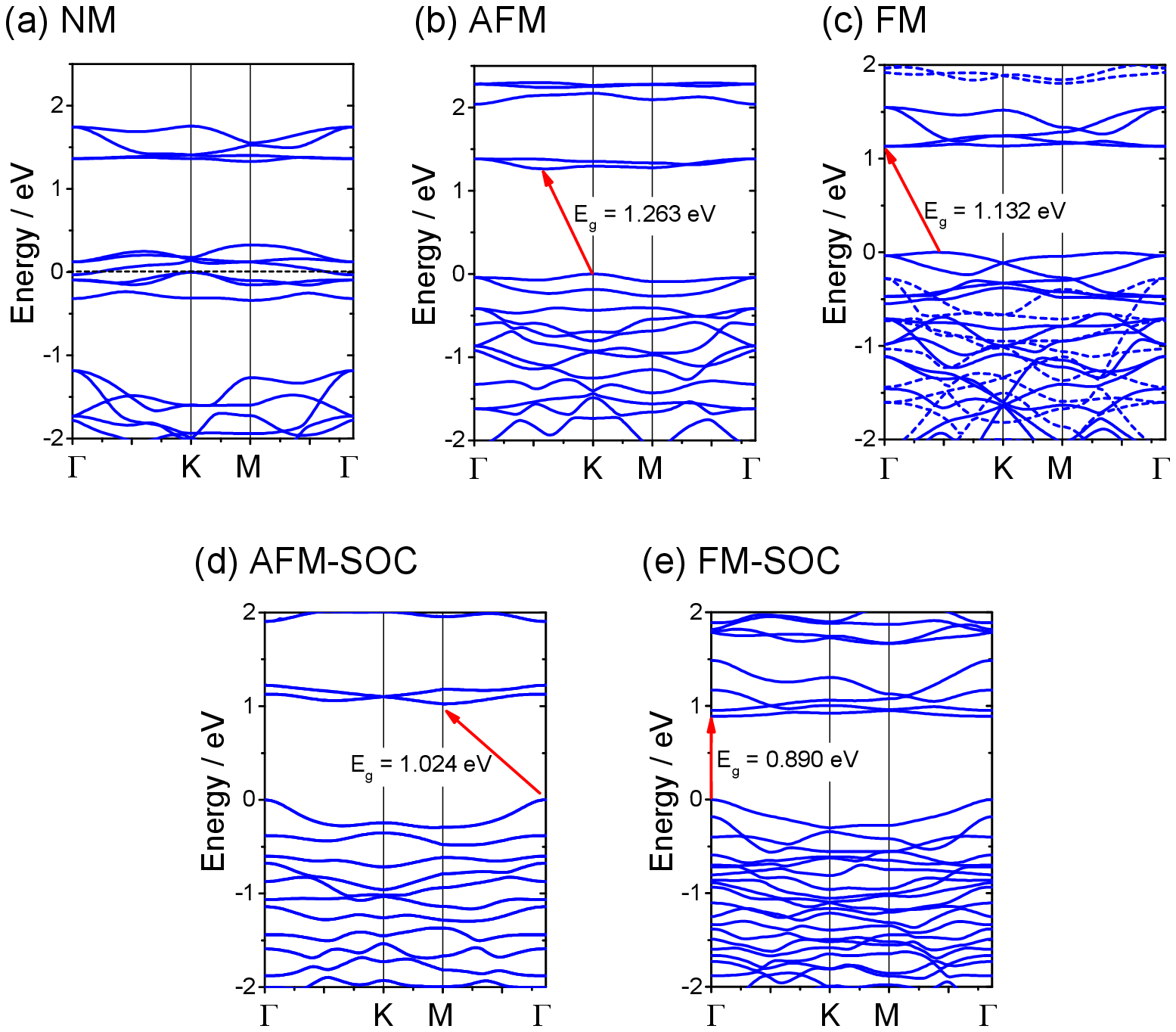}
\caption{\label{fig:suppband}Electronic band dispersions of monolayer CrI$_3$ calculated using LDA for the NM, AFM, and FM phases, respectively. (a) NM phase; (b) AFM phase without SOC; (c) FM phase without SOC; (d) AFM phase with SOC; (e) FM phase with SOC. In (b) and (c), solid lines denote spin-up while dashed lines denote spin-down bands. The energy band gaps between conduction band minimum (CBM) and valence band maximum (VBM) have been indicated using red arrows in each case. The Fermi level or the VBM has been shifted to zero.}
\end{figure}


\subsection{Lattice dynamics in PBE}
Listed in Table~\ref{supptab3} are the PBE results for the NM, AFM and FM phases, respectively. Unlike LDA, PBE yields non-negative phonon dispersions in monolayer CrI$_3$. We list these data for comparison.

\begin{table}[h]
\centering
 \caption{Frequencies of the vibrational modes at $\Gamma$ calculated using PBE with or without SOC for monolayer CrI$_3$ in NM, AFM, and FM phases, respectively. The Raman active modes have been highlighted in bold. The unit of frequency is in cm$^{-1}$. The symmetries of each mode in the FM phase have been indicated. } \label{supptab3}
\begin{ruledtabular}
\begin{tabular}{llllllc}
  & \multicolumn{1}{c}{NM}  & \multicolumn{2}{c}{AFM} & \multicolumn{2}{c}{FM} &  \\
    \cline{2-2}    \cline{3-4} \cline{5-6}
Mode No.   & PBE & PBE & PBE-SOC & PBE & PBE-SOC & Symmetry \\
   \hline
1, 2 & \textbf{27.8}	&	\textbf{45.9}  &	\textbf{45.1, 45.2} 	& \textbf{45.7}  &	\textbf{45.3, 45.4} & \textbf{$E_g$}  \\

3	 & 52.1 &   53.8  & 51.1	   	& 52.1  &	50.9       & $A_{2u}$ \\
4	 & \textbf{68.1} &	\textbf{68.0}  & \textbf{66.7}	    & \textbf{67.1}  & \textbf{66.7}       & \textbf{$A_{1g}$} \\

5, 6 & 84.4 &	73.6  & 72.8, 72.8	& 73.0  & 72.1, 72.2 & $E_u$ \\

7	 & 95.3 &	84.7  & 84.5		&	84.6  & 83.2  	  &  $A_{2g}$ \\

8, 9 & \textbf{87.3}     &	\textbf{93.1}  & \textbf{92.9, 92.9}	& \textbf{96.2}  & \textbf{94.8, 94.9}	&  \textbf{$E_g$} \\

10, 11 & \textbf{111.0}    &	\textbf{99.2}  & \textbf{97.8, 97.9}	& \textbf{98.9}	& \textbf{97.7, 98.0}	& \textbf{$E_g$} \\

12, 13  & 103.0 &	104.3 & 102.8, 102.9& 103.0	& 101.1, 101.5 &	$E_u$ \\

14	   & \textbf{125.9}   &	\textbf{117.2} & \textbf{115.1}		& \textbf{119.6}	& \textbf{117.6}	       & \textbf{$A_{1g}$} \\

15	   & 136.5   &	120.2 &	118.8		& 120.2	& 118.3	       & $A_{2u}$ \\
16	   & 147.4   &	200.1 & 196.6		& 196.6	& 193.0	       & $A_{2g}$ \\

17, 18  & 211.0  &	208.6 & 206.7, 206.8& 205.6	& 203.0, 203.1 & $E_u$ \\

19, 20  & \textbf{202.0}   &	\textbf{219.5} & \textbf{216.4, 216.4}& \textbf{221.0}	& \textbf{217.4, 217.9} & \textbf{$E_g$}\\
21	   &  225.6  &	241.8 & 239.1		& 244.8	& 241.0	       & $A_{1u}$ \\
\end{tabular}
\end{ruledtabular}
\end{table}

\begin{figure}[h]
\centering
\includegraphics[width=0.75\textwidth]{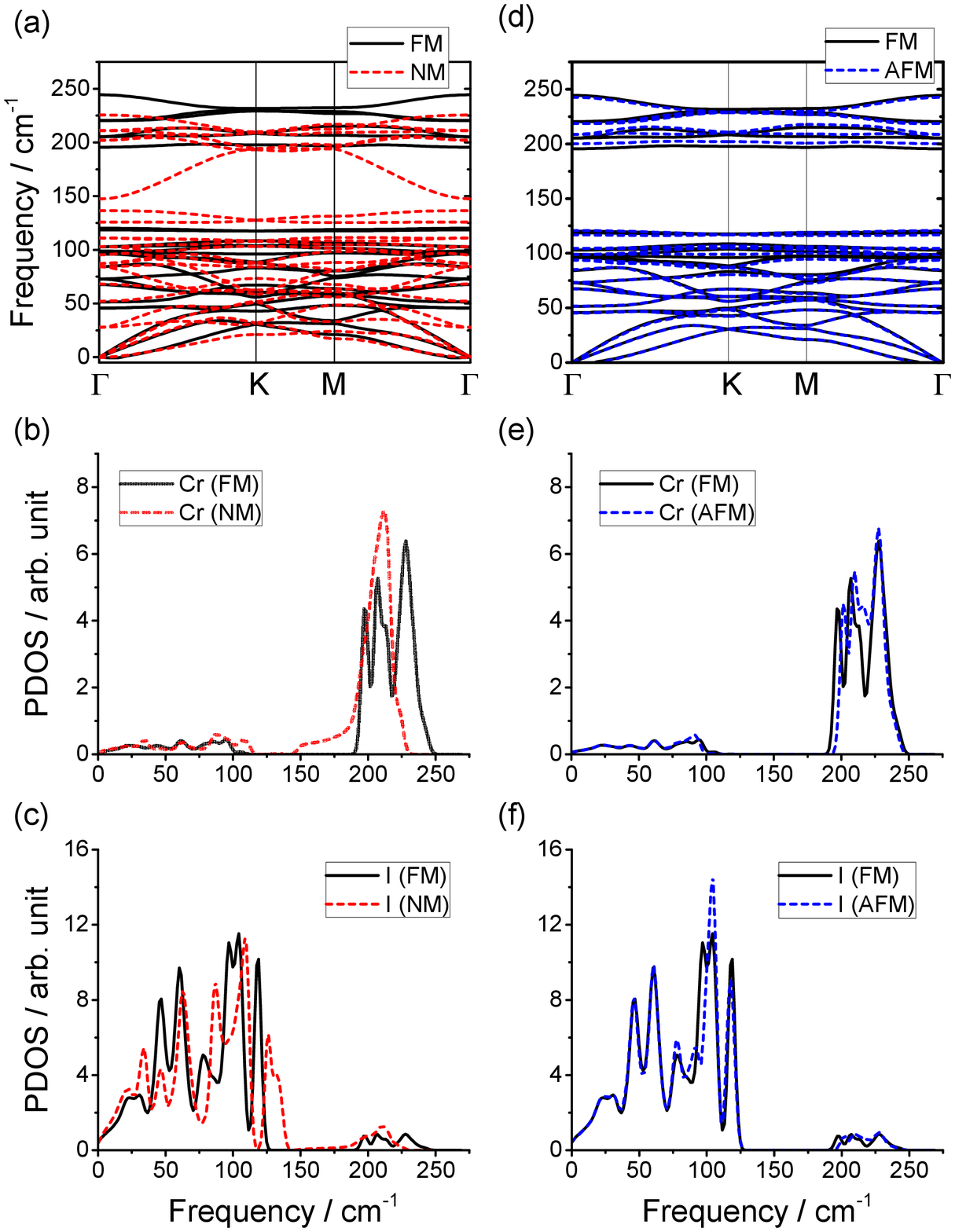}
\caption{\label{fig:suppphband}Phonon band dispersions of monolayer CrI$_3$ calculated using PBE for the NM, AFM and FM phases, respectively. Since FM is the ground state, we have used it as a reference. Left: (a) FM (solid) and NM (dashed); The corresponding partial phonon density of states are shown for two Cr atoms (b) and six I atoms (c), respectively. Right: (d) FM (solid) and AFM (dashed); The corresponding partial phonon density of states are shown for two Cr atoms (e) and six I atoms (f), respectively.}
\end{figure}

\subsection{Raman spectra in PBE}
The Raman spectra calculated in PBE are shown in Fig.~\ref{fig:suppraman}.
\begin{figure}[h]
\centering
\includegraphics[width=0.75\textwidth]{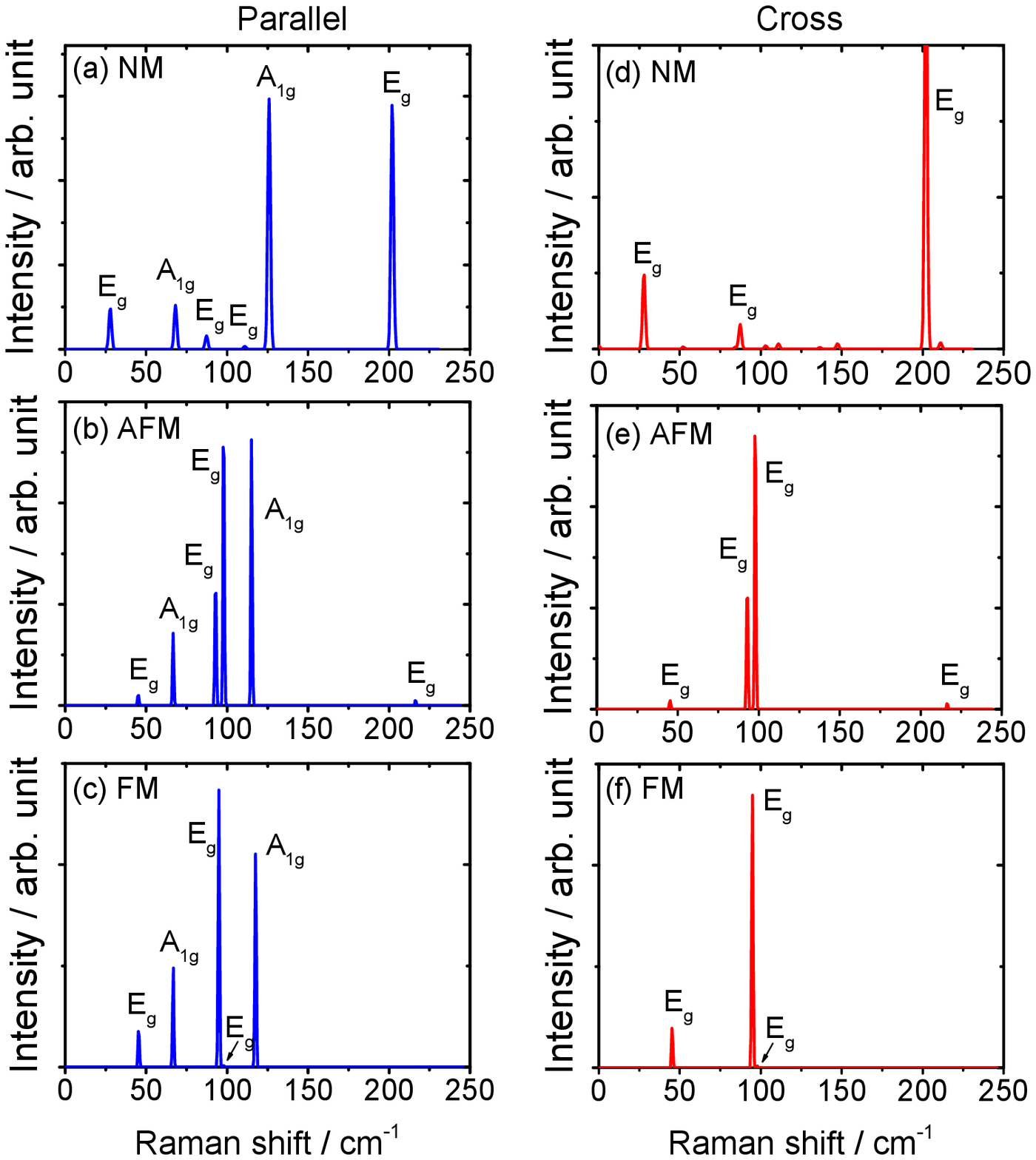}
\caption{\label{fig:suppraman}Polarized Raman spectra for monolayer CrI$_3$ in NM, AFM and FM phases, respectively. Note that since CrI$_3$ in the NM state is metallic, the dielectric tensor for Raman calculations was obtained at a typical experimental laser frequency 1.96 eV (633 nm). Shown on the left are for the parallel polarization laser set-up. On the right are the results with cross laser polarization set-ups. The corresponding mode symmetries for each peak have been indicated. A Gausian broadening with width of 1.0 cm$^{-1}$ has been applied.}
\end{figure}

}

\end{document}